# Understanding Technology Use in Global Virtual Teams: Research Methodologies and Methods

Tony Clear and Stephen G. MacDonell

*SERL, School of Comp. and Math. Sciences*
*Auckland University of Technology*
*Private Bag 92006, Auckland 1142, New Zealand*
*{tony.clear, stephen.macdonell}@aut.ac.nz*

## Abstract

***Context:*** The globalisation of activities associated with software development and use has introduced many challenges in practice, and also (therefore) many for research. While the predominant approach to research in software engineering has followed a positivist science model, this approach may be sub-optimal when addressing problems with a dominant social or cultural dimension, such as those frequently encountered when studying work practices in a globally distributed team setting. The investigation of such a team reported in this paper provides one example of an alternative approach to research in a global context, through a longitudinal interpretive field study seeking to understand how global virtual teams mediated the use of technology. The study involved a large collective of faculty and support staff plus student members based in the geographically and temporally distant locations of New Zealand, the United States of America and Sweden. ***Objective:*** Our focus in this paper is on the conduct of research in the context of global software activities, and in particular, as applied to the actions and interactions of global virtual teams. We consider the appropriateness of various methodologies and methods in enabling such issues to be addressed. ***Method:*** We describe how we undertook a substantial field study of global virtual teams, and highlight how the adopted structuration theory, action research and grounded theory methodologies applied to the analysis of email data, enabled us to deliver effectively against our goals. ***Results:*** We believe that the approach taken suited a research context in which situated practices were occurring over time in a highly complex domain, ensuring that our results were both strongly grounded and relevant to practice. It has resulted in the generation of substantive theory and techniques that have been adapted and applied on a pilot basis in further field settings. ***Conclusion:*** We conclude that globally distributed teamwork presents a complex context which demands new research approaches, beyond the limited set customarily applied by software engineering researchers. We advocate experimenting with different research methodologies and methods so that we have a more rounded repertoire to address the most important and relevant issues in global software development research, with the forms of rigour that suit the chosen approach.

**Abbreviations:** AIT - advanced information technology; AST - adaptive structuration theory; AUT – Auckland University of Technology; CTF – collaborative technology fit; Env - Environmental; GSD – global software development; GVT – global virtual team; LT – Local team; SE – software engineering; TUM – technology-use mediation; TUMAST – technology-use mediated AST; UMEA - User-Monitoring Environment for Activities.

## 1. INTRODUCTION

The globalisation of activities related to the production and use of software systems – from helpdesk offshoring through virtual infrastructure support to remote outsourced application development – has produced many significant challenges, opening up fertile ground for those interested in how those challenges can be addressed effectively. Many research questions arise: What version control techniques work in a global development context? Can collaborative technologies enhance global development productivity? How do individuals and groups relate across multiple cultures? What impact do time and space have on activity co-ordination? How does leadership manifest itself in dispersed teams? What role does the mediation of technology-use play in global virtual teams?

Evident in these questions is significant breadth of issues to be addressed, ranging from the largely technical – regarding techniques and tools – to the principally social – concerning culture and leadership. If software globalisation is to succeed, then *all* such issues need attention. And it is not a matter of 'one size fits all' when it comes to how these issues might be investigated – while questions with a technical emphasis may lend themselves to quasi-experimental analysis, those that are more social in nature are likely to require a very different research approach.

Our focus in this paper, then, is on issues of research methodology and method in the context of global software activities, and in particular, as they apply to the actions and interactions of global virtual teams (GVTs). Prior work that has applied differing research approaches is presented in the next section, drawn from both software engineering (SE) and information systems literature. We then consider the appropriateness of various methodologies in terms of enabling researchers to tackle research goals and answer questions associated with the actions and interactions of GVTs – addressed in Section 3. We also direct our attention to the applicability of specific research methods in this context. A longitudinal interpretive field study is then reported in Section 4 to demonstrate in detail the approach that we used in seeking to understand how global virtual teams mediated the use of technology. The study involved a large collective of faculty and support staff plus student members based in the geographically and temporally distant locations of New Zealand, the United States of America and Sweden. Finally we draw conclusions from our work in Section 5 and provide pointers to further research questions.

## 2. BACKGROUND AND RELATED WORK

In discussing approaches to assessing validity in the research process McGrath [36, p.14] has suggested that research:

> *"involves a) some content that is of interest b) some ideas that give meaning to that content, and c) some techniques or procedures by means of which those ideas and content can be studied".*

He terms these the substantive, conceptual and methodological domains, and then defines the research process as "the identification, selection, combination and use of the elements and relations from the substantive, conceptual and methodological domains." (p. 16). While there is much research related to Global Software Development (GSD) in both the substantive and the conceptual domains, the focus in this paper is on the methodological domain, and relevant research approaches are reviewed below.

### *2.1. Research Paradigms*

The underlying assumptions upon which a researcher may conduct an enquiry can differ markedly and provide a foundation for very different styles of research. As Dittrich et al. [16] note, qualitative research "may come in many different flavours…be used under different epistemological paradigms, and with different theoretical underpinnings". One useful categorisation of research paradigms positions them within three distinct approaches [41], each based upon a distinctive worldview and perspective on the nature of knowledge. From these originate three quite distinctive perspectives on the conduct of scientific enquiry, which Habermas [28] has depicted in a framework of "knowledge interests" presented in Table 1.

**Table 1.** The knowledge constituted interests of Habermas [28]

| **Interest** | **Knowledge** | **Medium** | **Science** |
|---|---|---|---|
| Technical | Instrumental (causal explanation) | Work | Empirical-analytic or natural sciences |
| Practical | Practical (understanding) | Language | Hermeneutic or 'interpretive' sciences |
| Emancipatory | Emancipatory (reflection) | Power | Critical sciences |

If we regard a research paradigm as a mechanism through which a researcher can assert the validity of particular truth claims, then we can view these as three distinct forms of truth supported by differing scientific approaches. For each of these belief systems a different research paradigm exists - the traditional or "classical" science 'objective' paradigm, the social sciences 'interpretive' paradigm, and the critical sciences 'evaluative' paradigm. Each paradigm comes with its own strengths and weaknesses, and as a result is better suited to answering particular research questions.

### *2.2. Research Methods in Global Software Development*

The study of global virtual teams and global software development has seen researchers contributing from differing traditions, with the Software Engineering and Information Systems disciplines contributing strongly to the extant literature. Software engineering has tended to favour the 'empirical-analytic' tradition of the natural sciences [2, 33]. Such is the extent of the use of experimental methods in software engineering that a systematic review of quasi-experimentation in software engineering research was reported by Kampenes et al. [31]. Studies utilising these and similar natural science methods such as surveys were shown by Glass et al. to be predominant in software engineering in a review of work reported in 2002 [24]. Information Systems researchers in contrast have moved towards a greater acceptance of research based upon the 'hermeneutic sciences', and the accompanying qualitative methods of the interpretive paradigm [34, 39].

While there does appear to be a growing acceptance of qualitative methods in software engineering, as reported in a recent special issue on "qualitative software engineering research", very different approaches may be taken in the conduct of qualitative research:

> *Qualitative research with a positivistic underpinning might be the most accessible one from a traditional software engineering background: qualitative researchers, like quantitative researchers, may present their conclusions about the data as objective, truthful statements about the world* [16].

Thus the researchers' epistemological stance is important in the design and conduct of research. However, as Glass et al. [23] have noted, there has been a tendency in academia for the hard – the technical – to drive out the soft – the behavioural. The increasingly prevalent challenge in software engineering research, however, and in global software development research as an illustration of this, is that such an approach may be inadequate when it is impossible to separate the software from the technology and, in turn, from the system and its human actors, their beliefs and perceptions. When isolation of the software as fits a reductionist research model is neither feasible nor tolerable, yet we need to arrive at insights that are both useful and defendable, the challenge is to adopt and become comfortable with new research methods. For instance, in a study investigating how virtual teams created "shared meaning" [6], the authors applied an "interpretive case study methodology", arguing that:

> *This methodology is appropriate because it focuses on the complexity of human sense-making in emerging situations and attempts to understand the phenomenon through the meanings that participants assign to actions and situations.*

In a later study building upon the work of Glass et al. [24], Segal et al. [47] classified the "research approach" of 46% of studies in the journal *Empirical Software Engineering* over the 1998 – 2003 period as "evaluative deductive" and only 2% "evaluative interpretive", noting that of all categories of research, "evaluative deductive – testing hypotheses in a very positivist tradition – dominated" [47]. While some 13% of papers were noted as applying a case study "research method", these appear to have taken a relatively limited approach to scientific evaluation. An apparently overlapping proportion applied a "research approach" termed "descriptive" (13% of papers) [47]. A ten year survey of the same journal from 1996 – 2006 more recently identified a slight shift in pattern of 37.6% experiments, and 28.6% case studies [30]. A recent broader study of computer science research indicated that the "engineering epistemology" of the discipline which largely involved "the proposal of new entities" led to an inbuilt bias against evaluation, with only 36% of the papers including a limited degree of evaluation of the "new entity" created [52].

An indicative analysis conducted to assess the research methods and approaches currently being used in global software development is presented in Table 2. This data is drawn from a subsample (arbitrarily the first 17) of research papers presented at the 2009 International Conference on Global Software Engineering (ICGSE 2009). The categorisation scheme for methods and approaches presented in [24] has been adopted, with the addition of a column for research paradigm to indicate the underlying epistemology for the work. While this analysis has been conducted by the first author only as a single rater and makes no claims to be exhaustive, it does offer a picture of research approaches quite divergent from those found in the studies of software engineering research in [24, 47, 30]. The relatively uncommon categories highlighted in bold in Table 2 were those most dominant in [24] and arguably in [52]. Therefore this suggests that the diversity of approaches is growing to deal with the broader range of issues to be addressed in a global SE context, and it may no longer be entirely accurate to state that "SE research is fundamentally about technical computing focused issues, and that it is seldom about behavioral issues" [24]. It is also evident from this brief snapshot that while most studies are conducted within an empirical epistemology some awareness is developing of the merits of interpretive studies.

**Table 2.** Favoured research approaches and methods in global software development

| **Paper No.** | **Research Approach** | **Research Method** | **Research Paradigm** (**E**mpirical/ **I**nterpretive) |
|---|---|---|---|
| 1 | Evaluative-deductive | Descriptive/exploratory survey | E |
| 2 | Evaluative-interpretive | Grounded Theory | I |
| 3 | Evaluative-interpretive | Grounded Theory | I |
| 4 | Descriptive other | Data analysis | E |
| 5 | Evaluative-other | Descriptive/exploratory survey | E |
| 6 | Evaluative-deductive | Descriptive/exploratory survey | E |
| 7 | Evaluative-deductive | Descriptive/exploratory survey | E |
| 8 | Evaluative-other | Descriptive/exploratory survey | E |
| 9 | Descriptive system | ***Concept implementation (proof of concept)*** | E |
| 10 | ***Formulative-framework*** | Laboratory experiment (software) | E |
| 11 | Evaluative-interpretive | Grounded Theory | I |
| 12 | Descriptive other | Descriptive/exploratory survey | E |
| 13 | Evaluative-deductive | Descriptive/exploratory survey | E |
| 14 | Evaluative-deductive | Instrument development | E |
| 15 | Review of literature | ***Conceptual analysis*** | E |
| 16 | Evaluative-deductive | Case study | E |
| 17 | Evaluative-deductive | Data analysis | E |

## *2.3. Choice of Research Paradigm to fit the Research Goals and Questions*

In the field study profiled in this paper (further elaborated in Section 4) the phenomenon being investigated was that of technology-use mediation in global virtual teams [12]. Technology-use mediation (TUM) as proposed by Orlikowski et al. [43] refers to activities undertaken by those involved in *supporting the use of* information technology, rather than directly *using* the technology itself. For GVTs this involves the work of personnel who play intermediary roles that significantly impact upon GVT outcomes, yet are not well understood: "Technology facilitation has been an important, yet neglected topic for many years…about which we know little. Still, its importance seems to have increased as work has become increasingly computer mediated." [50, p.85].

In deciding how to go about the research, several questions presented themselves. The phenomenon of TUM in a global collaborative context encompassed a diverse set of roles and activities, and its operation could only be explored as it unfolded over time. Thus a snapshot-based research method such as a survey (e.g. [5]) was not appropriate. Likewise since the phenomenon had been largely "neglected" [50] we lacked an established theoretical grounding for the research and were therefore not in a strong position to formulate hypotheses and conduct confirmatory experiments.

There were also significant differences between the three-site relatively ad hoc collaboration engaged in here and the "follow the sun" [51] and tightly structured three-location "24 hour factory" models outlined in [27, 48]. Here we had a more fluid and heterogeneous situation with peer organizations collaborating in a relatively loose and informal manner. It exemplified both an intrapreneurial and extrapreneurial collaborative venture, without formal contracts or official sanction at organization levels, and individuals with a variety of roles so that the "composite persona" notion, where the coordinators at each of the three sites acted as in effect one person [26], was hard to put into effect even for the academics involved. Students and their teams of course were much more unbiddable! In that sense this collaborative model was more analogous to a small enterprise situation and not the large corporate environment. Characteristics of the setting included: a heterogeneous infrastructure; ad hoc peer collaboration; loosely negotiated lead parties; no opportunity for the coordinators to meet face to face; and no specific software and infrastructure support for global collaboration processes (e.g. tailored project event tracking systems such as UMEA or Multimind [32, 26]).

Another crucial set of research issues that needed to be addressed were those of data availability and accessibility. What data were required to investigate the phenomenon and what data could be made available? Then, having resolved those questions, what issues would present themselves in the pragmatics of data analysis?

A partial set of answers came from the wider programme of research into global virtual teams and global collaboration within which this study was embedded [13]. As an ongoing programme of action research [3, 4, 37] this provided the context within which the data were gathered. However, at the time of the global collaborative cycle reviewed in this study, the model of interactions was very much exploratory rather than confirmatory, with multiple emergent research goals (described in Section 3). The chosen model of action research is developed further in Section 3.2 below, but the choice of this research method had certain resulting impacts. Action research "qua definition is not controllable" [16], and therefore brings with it some inherent assumptions that require the adoption of qualitative research methods [16]. Commonly "qualitative analytical techniques like hermeneutics, deconstruction and theoretical sampling" [3] accompany action research, and Baskerville has asserted that:

> *"since action researchers adopt interpretive and ideographic postures they must also adopt qualitative data as a medium to the empirics"[3].*

An interpretivist perspective is not wholly fundamental to action research, however, with it being observed that:

> *"the underlying philosophy shared by most forms of action research is pragmatism. As a philosophy pragmatism concentrates on asking the right questions and getting empirical answers to those questions. On its own it does not explain very much, but provides a method to help explain why things work (or why they do not work)" [4].*

The theoretical basis of this study also lies in a structurational perspective as noted in Section 3.3 below. Speaking generally, Poole and DeSanctis [46] have observed that as a meta-theory "structuration theory leaves decisions about research settings, procedures, measurements and analytic tools to the researchers themselves". Structuration Theory [20] has been cited as an "integrating meta-theory" [42] that aims to reconcile the opposing empirical and interpretive philosophical perspectives within sociology, and thus could be viewed as a philosophical paradigm in its own right. Conducting empirical analysis then, with a coherent and robust set of tools and techniques, has required a suitable methodological 'toolkit' to be devised. While this study, in the pragmatic spirit of action research, therefore combines both perspectives with a strongly data-grounded empirical base, it would be more likely to be viewed by a software engineering researcher as an interpretive field study. Thus both the topic of interest – global virtual teams – and the embedding context – action research – have brought implications relating to the underlying epistemology, and for choice of research method and accompanying analytical techniques.

# 3. METHODOLOGIES AND METHODS FOR STUDYING GLOBAL VIRTUAL TEAMS

In this section we outline the goals we sought to achieve in our field study, the methodologies and methods used to address them, and the nature of the data we had at our disposal.

## *3.1. Research Goals*

There were three primary research goals in this study, which addressed the *substantive*, *conceptual* and *methodological* domains of McGrath [36] respectively.

- Firstly to investigate the role of 'technology-use mediation' in supporting the work of global virtual teams (GVTs).
- Secondly to develop and apply a framework for researching technology-use mediation in global virtual teams.
- Thirdly to gain deeper insight, in order to develop frameworks for the guidance of researchers investigating global virtual teams.

## *3.2. Research Methodologies and Methods Chosen*

Given the nature of the research topic, the limited initial level of understanding of the phenomenon, and the exploratory and "explanatory" [25, p.624] intent of the research, it did not seem appropriate to define a more tightly focused set of goals from the outset. The adopted set of research goals, in the words of Gregor [25, p.624], aimed at developing a "theory for understanding…how and why things happen in some particular real world situation". In contrast, much work in software engineering maps to another of Gregor's models in developing theories for predicting (cf. Glass et al. [24, 23]). Such a theory "says what is and will be…provides predictions and has testable propositions" [25, p.620].

From its origins this study had been grounded in the 'real world', with resulting theory deriving from that setting. In the McKay and Marshall [37] variant of action research, the separate components of the research are identified and consciously addressed. Five elements are noted within their framework, which enable a conscious separation of the practice components from the research elements, and thus enable the research to avoid the trap common to action researchers of having their work described as simply "consultancy". These five separate elements comprise:

1) [F] the research framework or conceptual element informing the research;
2) [$M_R$] the research method to be adopted;
3) [$M_{PS}$] the problem solving method that will be used in the practice situation;
4) [A] the problem situation of interest to the researcher (the research questions);
5) [P] the problem situation in which we are intervening (the practice questions of interest to the practitioners).

The elements of the action research framework as planned at the initiation of the study iteration in focus here are given in Table 3. These five elements in combination provided a concise means of planning and framing the research endeavour, capturing both the research and practice dimensions. The practitioner interest here related among other things to "improving the viability of student or software teams engaged in international teamwork". Technology-use mediation (TUM) was considered a key dimension in supporting the work of global teams in both contexts.

While the framework has the benefit of separating out the researcher and practitioner interests into distinct elements, it is unspecific about the underlying epistemology or ontology. Action research, as with all research, is conducted with some basic assumptions. These assumptions may be usefully aligned with the three main research paradigms identified in Section 2, being the 'empirical-analytic', the 'interpretive' and the 'critical'. Carr and Kemmis [9, p.136] echo this perspective with three variants of action research each grounded in a distinct worldview:

- *technical action research*, where the researcher acts as the expert and agenda setter, guiding a practice community towards some change based upon that agenda.
- *practical action research,* where the researcher acts more in the role of a process facilitator and is conducted in collaboration with the community towards joint goals.
- *emancipatory action research,* where the researcher and the practice community unite to address distortions and power imbalances in their situation.

**Table 3.** Elements of research investigating TUM in GVTs within a 'dual cycle action research' framework

| **Element** | **Description** |
|---|---|
| **F** (Framework) | • Extended Adaptive Structuration Theory (Clear, 1999 [11] & revisions in progress) |
| **$M_R$** (Research Method) | • Practical Action Research, with some aspects of emancipatory action research.<br>• Content analysis of online data (email, discussion threads, websites, Notes forms etc.) will incorporate grounded theory for TUM elements |
| **$M_{PS}$** (Problem solving method) | • Prototyping<br>• Collaborative Trials<br>• Practical Action Research<br>• Reflective practitioner model |
| **A** (Problem situation of interest to the researcher) | • How does TUM operate and support or hinder the work of GVTs?<br>• How does TUM operate and support or hinder e-collaboration?<br>• How do TUM, e-collaboration and GVTs interrelate? |
| **P** (Problem situation in which we are intervening) | • Improving teaching & learning through active learning approaches<br>  • Students as active co-researchers<br>  • Collaborative learning models<br>• Developing student capabilities in teamwork, cross cultural communication and use of IT<br>• Providing an interesting & meaningful learning experience<br>• Using e-collaboration to teach and practically demonstrate key concepts of groupware and group decision support<br>• Improving viability of student or software teams engaged in international teamwork |

The action research variant adopted here of "practical action research" is based upon a largely interpretivist and pragmatic worldview (notwithstanding the observations made relating to a structurational perspective in Section 2.3 above), in which the activities in co-operation with practitioners are mainly concerned with improving practice, and encouraging professional reflection.

### *3.3. Theoretical Underpinning*

Although there is extensive pragmatism in the McKay and Marshall [37] 'dual cycle' model of action research, encompassing both researcher and practitioner interests, the field study considered here was soundly informed by a theoretical underpinning. As indicated in Table 3, the theoretical framework informing the study was an extension of the "Adaptive Structuration Theory" (AST) of DeSanctis and Poole [15]. The AST framework was an 'input-process-output' framework initially conceived to guide research into Group Decision Support Systems.

The framework accommodates the *input* conditions of technology, environmental, task, group structure and group dynamics, enables study of the structuring *processes* of technology appropriation, and assesses *outputs* (generally the outcomes) of the group decision process.

Prior research studies of these group support or 'electronic meeting systems' have mostly investigated the work of co-located and synchronous groups. In contrast, global virtual teams carry out projects with members who span country, time-zone and institutional boundaries, may have never met face-to-face, and communicate primarily through technology-supported modes. Asynchronous forms of communication are common and for these teams to effectively function, a cast of supporting actors must work actively behind the scenes, to mediate their information technology use. By that we mean such activities as *establishing the technology*, and ongoing processes of active *reinforcement* and *adjustment* to embed productive patterns of collaborative technology use, interspersed by instances of significant *episodic change*. The extensions to AST referred to in Table 4 were intended to accommodate this dimension of technology-use mediation by incorporating these four primary activities of TUM.

To illustrate TUM, an example of the four different types of mediating activities carried out by a network administration group were identified in a study by Orlikowski et al. [43]:

1) **establishment**: established role, determined and built consensus around use of the communication technology, established guidelines etc. for its use;
2) **reinforcement**: training, monitoring, and follow-up with members and the group to reinforce the established guidelines;
3) **adjustment**: on the basis of feedback obtained from members, adjusted the definitions and usage rules for specific newsgroups and occasionally added new newsgroups on request;
4) **episodic change**: twice during the project, the network administration group initiated major changes to the news system as a whole.

In the course of the study reported here (and in greater detail in [12]) a novel unifying framework known as "Technology-use Mediated Adaptive Structuration Theory" (TUMAST) was developed. It was conceived specifically to address the second research goal stated above, namely, *to develop and apply a framework for researching technology-use mediation in global virtual teams*. Thus it served a methodological role in enabling the research to be conducted. The propositions for AST [15] and the TUMAST extensions are briefly presented in Table 4, with an accompanying schematic for TUMAST in Figure 1. For a fuller exposition of the relevant notions the reader is referred to [15] and [12].

**Table 4.** Propositions of AST and TUMAST (expansion upon [15, p. 128ff.])

**AST**

**P1.** AITs (Advanced Information Technologies) provide social structures that can be described in terms of their features and spirit. To the extent that AITs vary in their spirit and structural feature sets, different forms of social interaction are encouraged by the technology.

**P2.** Use of AIT structures may vary depending on the task, the environment, and other contingencies that offer alternative sources of social structures.

**P3.** New sources of structure emerge as the technology, task and environmental structures are applied during the course of social interaction.

**P4.** New social structures emerge in group interaction as the rules and resources of an AIT are appropriated in a given context and then reproduced in group interaction over time.

**P5.** Group decision processes will vary depending on the nature of AIT appropriations.

**P6.** The nature of AIT appropriations will vary depending on the group's internal system.

**P7.** Given AIT and other sources of social structure, n1 ….nk, and ideal appropriation processes, and decision processes that fit the task at hand, then desired outcomes of AIT use will result.

**TUMAST**

**P8.** The activities of technology-use mediators offer an "alternative source of social structures" for P2 above.

**P9.** Technology-use mediators are instrumental in P2's "use of AIT and social structures", through the TUM activities of 'establishment' and 'reinforcement'.

**P10.** Technology-use mediators are instrumental in P3's "emergence of new sources of structure", through the TUM activities of 'adjustment' and 'reinforcement'.

**P11.** Technology-use mediators are instrumental in P4's "emergence of new social structures", through the TUM activities of 'episodic change'.

As an extension of the prior AST model the three constructs dealing with *sources* and *forms of structure* were augmented, by adding the roles of *technology-use mediator*s, with the assumption that appropriate technology-use mediation (TUM) related activities would be conducted during those input, process or output stages. This model enabled the analysis of TUM activities by situating them within discrete windows of activity in a context of collaborative technology use in a global setting. These activity windows could be analysed one by one or in combination by building them up in input-process-output-input sequences, as might be encountered within a multi-phase project. The usefulness of this TUMAST framework for investigating the phenomena associated with GVTs and TUM has been tested, through its application to the work of global virtual teams, and those supporting them, in one particular six-month research iteration as reported here, and in full in [12].

In their review of the application of qualitative research strategies in software engineering, Dittrich and colleagues [16] reported studies employing methodologies such as action research, structuration theory and grounded theory. In the study reported here these strategies were applied not singly but in combination. Action research provided the overarching methodological framework within a wider programme of research, under whose umbrella this study into TUM comprises one span. Structuration theory informed the AST and TUMAST theoretical models, applied within the study, and the subsequent process of analysing the data. Then further, Grounded Theory [22] was applied. This combined approach was adopted in order to develop a "theory for explaining" [25, p.624] the complex and recursive phenomenon of TUM in a GVT context. The richness of this combination of approaches also enabled the longitudinal study of a complex set of activities and a series of actions as they evolved.

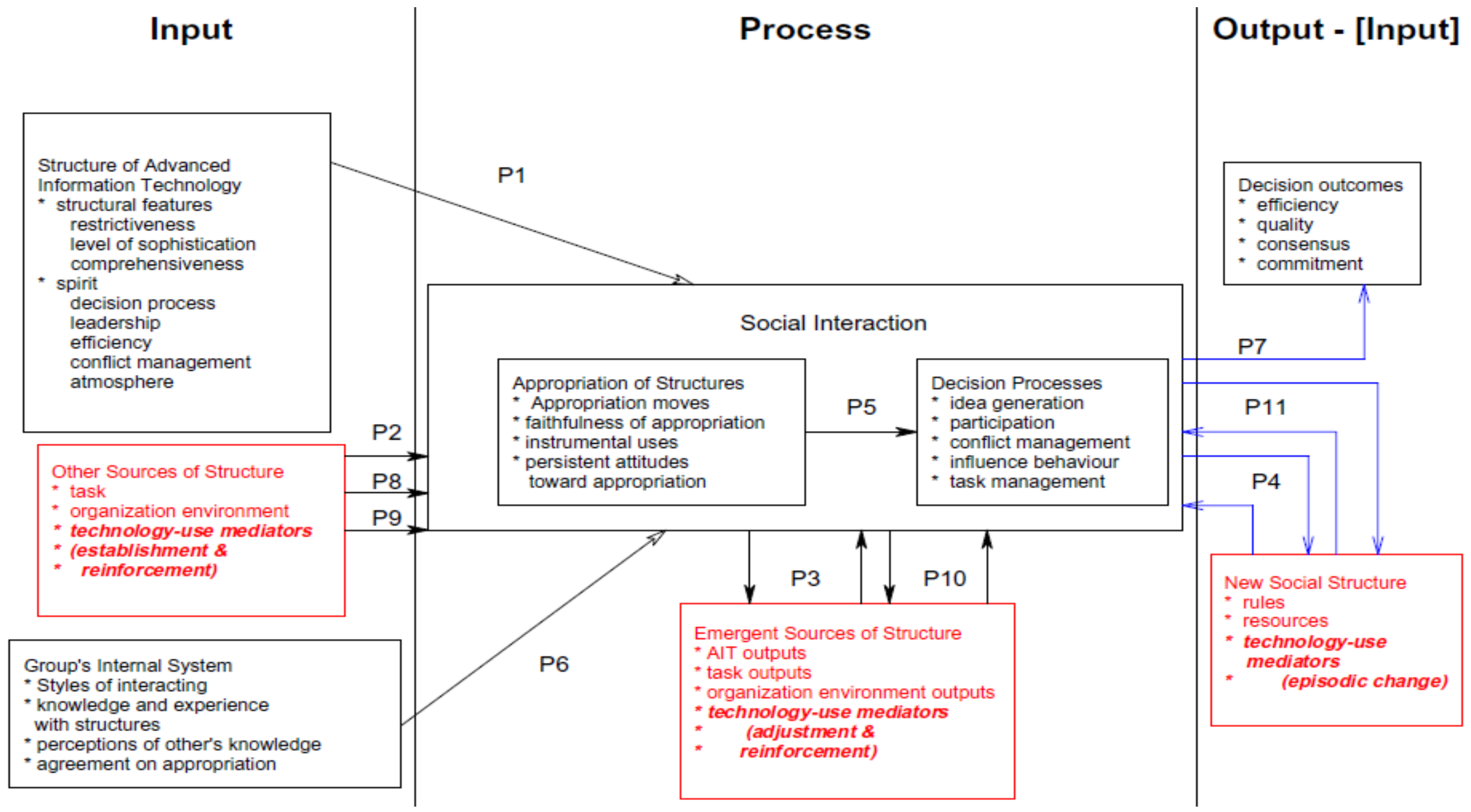

**Figure 1.** The Technology-use Mediated AST Framework (TUMAST) adapted from AST [15, p.132] & [11]

### *3.4. Characteristics of Data Involved*

Defining the meaning of data in action research projects, especially in a complex domain such as global software development, can be problematic, with historical, contextual, process, empirical and evaluative forms of data [10, p.111] all contributing to a rich mine for analysis.

The data of primary relevance in this project was the 'empirical' data, represented in particular by the email messages exchanged between the actors in the collaboration. This formed a corpus of email data from multiple contributors, spanning a period of more than one year's duration. The corpus incorporated data originating from or going to some 20 participants in the study (including the first author as project coordinator). Participant roles were diverse and included: trial coordinator, on-site coordinator, researcher, class room educator, student, Lotus Notes administrator, network administrator, developer, system support consultant, help desk staff, resource coordinator, ethics committee, audio-visual unit, Flexible learning manager. The study had further electronic and hard copy data sources – online postings, discussion threads, Lotus Notes database entries, evaluation forms, shared documents and files, literature entries, diary notes, hard copy folders and others. Managing this diversity of sources itself presented a challenge. As Allan [1] has observed, in grounded theory research "data collection is usually but not exclusively by interviews", so we needed to experiment in this global context in order to develop new methods and a data collection and coding approach that would handle email data as the primary source of data for this study. Thus the analysis, although interpretive, would be firmly grounded in the data and strongly empirical by nature.

Space precludes a full discussion here of the challenges encountered in developing methods for managing and coding the large amounts of data within the study. However the process of segmenting the data for analysis, and defining units of analysis at differing levels, does warrant mention. The input-process-output models underlying AST and TUMAST and the notion of a process model with antecedent conditions, a sequence of events comprising development and resultant outcomes, led to the identification of an *episode* as a logical unit of analysis. The notion of an individual episode or an *"episode of interest"* as an analytical unit in the study was defined as:

> *A relevant temporally bound sequence of events with antecedent conditions and outcomes, which stands apart from others, and has been selected for analysis.*

Criteria for relevance included such considerations as: Does the episode present a specific example of TUM activity? Or, does it exemplify one or more of the TUM modes? Short episodes may be selected on the basis of some form of critical incident in which TUM activity is notable. Typically, such incidents presented themselves as some form of "breakdown" [29] in the collaboration process. Alternatively, longer episodes might be selected by a logical time-bound unit, in this case the full duration of the *establishment* TUM phase for the cycle of global collaboration, where the TUM mode determined the temporal boundaries.

Examples of selected episodes are presented in Sections 4.1 ff. below. As can be seen in those examples the temporal delineation is a key element of an episode. While an "episode of interest" may be summarised by an accompanying narrative, it is not a "story" with an authorial voice and a moral, amenable to analysis via narrative enquiry. Pentland for example has categorised electronic mail logs as "annals" rather than narratives [44]. Likewise it is not a scenario or a use case [54] as might be modelled in a software engineering context, as it is not intended to lead to design, but it depicts a specific set of events, communications and actions. To date these have been retrospective, but in concept a future scenario of technology use (e.g. establishing a shared project management and wiki platform [7] as communication media for a project team) might be envisioned as an episode, if we wished to predict the outcomes of that technology implementation. An episode may incorporate discourse in the form of a sequence of communications and perhaps be amenable to various forms of discourse analysis [49], but it will typically also embody individual actions, institutional, cultural and technology dimensions. Thus it differs from these other analytical units.

Representative temporally bound episodes were therefore selected for analysis based upon a "theoretical sampling" strategy [22], within the overall TUMAST framework of the four modes of TUM activity (*establishment, adjustment, reinforcement* and *episodic change*). An episode needed to exemplify the TUM activity for the chosen mode. As noted, the full *establishment* phase was chosen as a logical temporal unit, to analyse the *establishment* mode of TUM in the global exercise. For the other phases the strategy involved selecting relevant "breakdowns" [29, 53] as 'critical incidents' in which the technology had moved from the background to the foreground, and become 'unconcealed'. These incidents provided notable occasions for reflection and TUM activity, most obviously for the *adjustment* and *reinforcement* modes. Episodes in the *episodic change* mode were again selected on the basis of a longer term response to a 'breakdown' incident, or as an evident juncture in the flow of the team's collaborative activity.

The episodes presented themselves progressively as the data were prepared for analysis, a further indication of the emergent nature of the processing of data in an interpretive study such as this. This enabled the achievement of a justifiable and manageable theoretical sampling strategy [22] to select the data and define it within relevant episodes, which would support this investigation of TUM in GVTs. We believe that the process of selecting data is replicable, following these criteria, although the precise episodes might differ in any selection process. Thus the approach enables the application of a "replication logic" [17], through the analysis of multiple 'cases', where each case may present confirming or disconfirming patterns. Given the adopted "theoretical sampling" mode and the desire to demonstrate "theoretical saturation" [22], further episodes could thus serve a confirming or disconfirming purpose.

The eight episodes eventually chosen were selected in order to represent typical aspects of each of the four TUM activities (*establishment, adjustment, reinforcement* and *episodic change*) in operation, and to support their comparison across episodes. The *establishment* activity was represented by one full length episode; *adjustment-reinforcement* (typically in combination) by four smaller episodes; and *episodic change* by three episodes. The grounded theoretic method of data analysis adopted required that data analysis be conducted using the "constant comparative method" [21] whereby "while coding an incident for a category compare it with the previous incidents in the same and different groups coded in the same category" [22, p.106]. This constant comparison continued only until "theoretical saturation" [22] for a category had been reached, and we believe that the diversity and quantity of data across this set of episodes supports such analysis.

# 4. FIELD STUDY ANALYSIS AND INSIGHTS

This particular study of technology-use mediation in global virtual teams occurred in the context of a long term action research programme into global virtual teams (GVTs), collaborative computing and international collaboration (cf. [13]). Annual global virtual collaborations between undergraduate business students majoring in information technology in New Zealand and computer science students in Sweden have been conducted since 1998. The collaboration reviewed here therefore constitutes one cycle of many within the wider research programme. Participants in the global collaboration considered here were students from AUT University in Auckland New Zealand, from Uppsala University in Sweden, and from St Louis University Missouri in the United States of America. Students were formed into nine global virtual teams to perform the collaboration, during which they aimed to jointly complete a common decision making task. The overall task design for student GVTs comprised a framework of three primary elements: an icebreaker task; a collaborative task; and an evaluation and individual report. The supporting technology platforms comprised a combination of: AUTonline (the AUT virtual learning environment (based upon the commercial Blackboard product)), which included features such as group discussion forums, chat and file sharing capabilities; email; and a custom developed Lotus Notes database with a set of online forms for such activities as confirming group leaders and conducting online evaluation questionnaires.

The activities undertaken by an extended cast of supporting actors, who performed a variety of technology-use mediating roles to enable the exercise, were those of a further and distinct global virtual team working in a naturalistic and challenging professional context. Members of this team had links to other groups both within their own organizations and across organizational boundaries, as illustrated in Figure 2. The unfolding of their activities over time (rather than those of the student GVTs) is the focus of this study.

In order to investigate these activities the study concentrated upon the set of selected episodes that exemplified TUM activity as noted in Section 3.4. Each of the eight distinct episodes was analysed in depth by applying a standard pattern which contained six separate steps. Selected episodes are profiled below to indicate how the process of analysis was carried out. The full four modes of TUM activity are addressed in the episodes shown. To demonstrate the process of analysis the initial *establishment* episode is covered in full, followed by brief summaries of two other episodes respectively representing the *adjustment/reinforcement* and *episodic change* modes.

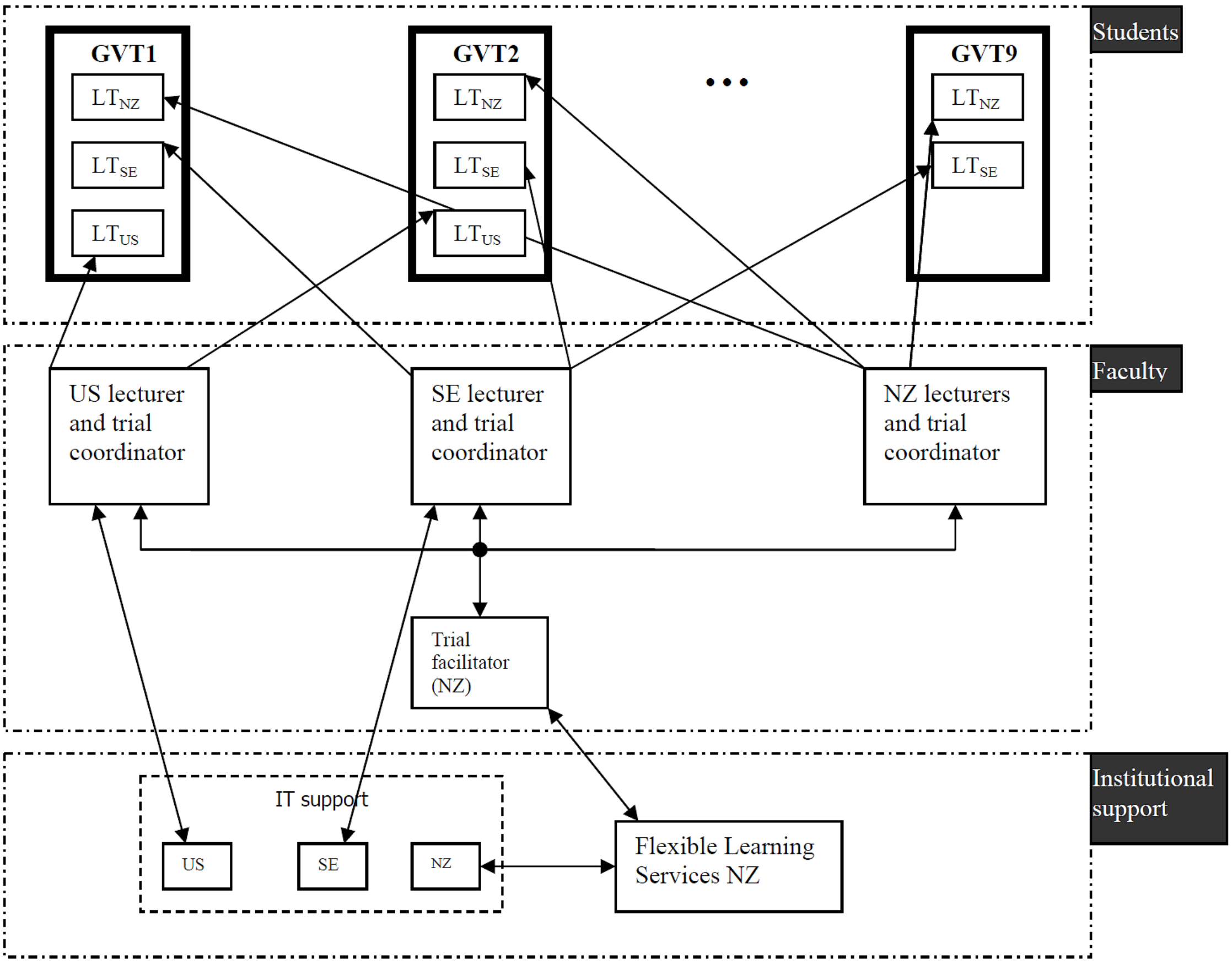

**Figure 2.** Collaboration process and roles of those involved[1]

### *4.1. Episodic Analysis - Establishment Episode*

For this major episode focusing on the *establishment* mode of TUM activity, **firstly** an overview of the episode was tabulated, indicating the broad scope of the episode in terms both of actors involved and data sources. This overview is shown in Table 5.

**Next** a narrative summary of the events was provided as briefly excerpted below.

> *…This episode consists of the full window of establishment TUM activities selected prior to the collaboration. The prior episodes have been in the nature of 'micro' episodes being based upon relatively few source items, but this episode (as can be seen from table [5] above), draws upon the work of numerous actors and a varied set of data items. With 216 source items, 15 actors, some third of a million words and expanding over a full year's duration, this analysis window could properly be termed a 'macro' episode. The end boundary of the establishment phase for the collaboration has here been set at 17/09/2004, the date at which the trial effectively started, despite 6/09/2004 having been planned as the official start date. Exceptions to that boundary are two student communications on 4/10/2004 and 14/10/2004 querying the state of play, and indicating that their groups had not started yet, suggesting that the establishment process for them at least had not yet taken effect.*
>
> *The episode includes a rich sequence of interconnected TUM activities, which together serve to establish the conditions within which the planned student GVT's are to function…*

The **third** element of *appropriation analysis*, [15, p.135], traversed four groups of 'appropriation moves' reflecting ways in which technology structures were appropriated (*direct; constraint; relate* and *judgement*). Figure 3

[1] Note: Figure 2 originated from the work of Diana Kassabova, a colleague in the 2004 trial.

presents the graph for ‘constraint’ moves in the episode, followed by the accompanying analysis.

**Table 5.** Episode characteristics - establishment episode

| Episode Characteristics | | |
|---|---|---|
| Duration: | 04/09/2003 –14/10/2004 | |
| Supporting data: | No. | |
| | 1 | Email Message: AB 16/09/2004 |
| | 1 | File: AB 16/09/2004 |
| | 22 | Email Messages: AP 23/06/2004 – 17/09/2004 |
| | 3 | Email Messages: BB 24/08/2004 – 16/09/2004 |
| | 5 | Email Messages: BD 30/06/2004 – 23/08/2004 |
| | 80 | Email Messages: DK 23/06/2004 – 14/10/2004 |
| | 10 | Files: DK 18/08/2004 – 17/09/2004 |
| | 1 | Email Message: A Pseudonym 01/07/2004 |
| | 38 | Email Messages: FN 10/06/2004 – 14/09/2004 |
| | 6 | Files: FN 05/09/2003 – 01/09/2004 |
| | nn | Etc…….. |
| No of sources | 216 | |
| Word count | 367, 973 | |
| Actors: | 15 | AB, AP, BB, BD, DK, APs, FN, FT, F, GG, KK, MD, MN, NI, TC |

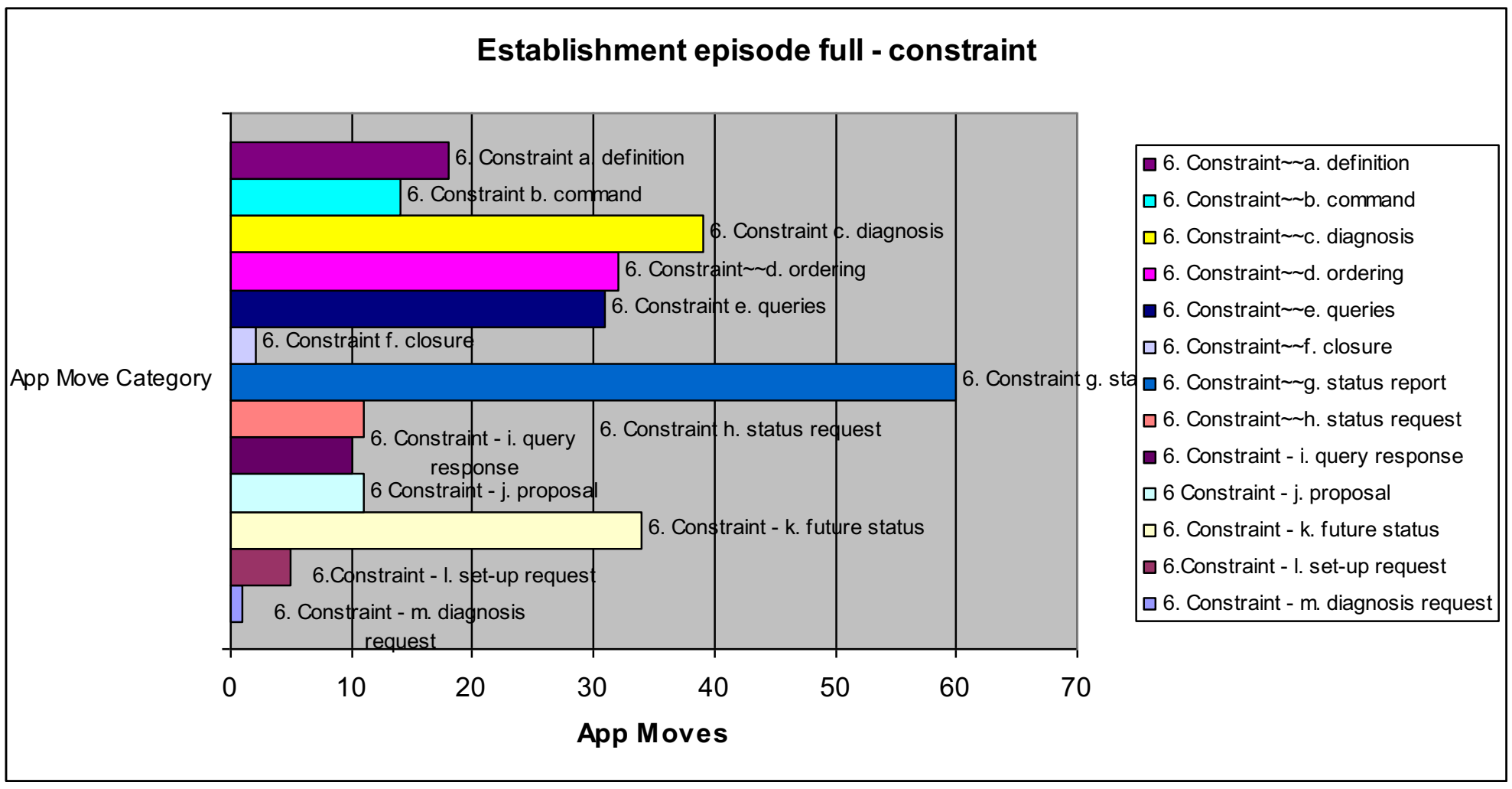

**Figure 3.** Constraint appropriation moves for establishment episode

Examples in the … subtype of “diagnosis” reflect “comments on how the structure is working” (positive or negative). In this grouping we see an ongoing interchange about creating and sharing the metastructure of “student lists” between sites, for entering into the technology platform represented by AUTonline. AP noted that:

> Students have the right not to give out their email addresses, and so there are some students for whom I do not have contact addresses yet. (AP 09/09)

With DK responding:

> As far as I understand the process here, emails are needed so the students can get their login info by email. (DK 08/09)

The **fourth** element of the analysis tabulated the grounded theoretic concepts and codes arising from the episode. Some 100 discrete codes were identified within this large episode. A subselection identifying merely the roles in evidence is portrayed in Figure 4 below.

The **fifth** element consisted of a set of visual mappings of selected aspects of the episode. This mapping applied radar charts to depict the operation of selected *metastructures* (mediating institutional, cultural or technology structures, which served to shape technology use - a concept developed in the course of the study and based upon the insights drawn from the prior grounded analysis). Space precludes a comprehensive discussion of this portrayal here, but these ‘visual maps’, such as those shown in Figure 5, supported by descriptive summary tabulations for each of the six dimensions (as shown in Table 6), were used to portray at a glance multidimensional aspects of the episode, and frequently served to isolate significant and problematic variation across the three sites. Table 6 and Figure 5 specifically portray the TUM activity associated with the process of forming the GVTs for the collaboration, and serve to illustrate how this form of analysis was conducted.

| **establishment episode full** | Role | socio-emotional group-bldg and mtce roles | Motivator (energizer, encourager) | Team leaders or session owners | Explainer (elaborator, coordinator, orienter, summarizer, amplifier) | Innovator | Formal (teaching~research assistants) | Purpose agents - teacher |
|---|---|---|---|---|---|---|---|---|
| | 0 | 4 | 69 | 1 | 1 | 1 | 6 | 44 |

| Researcher | Undergraduate Student | curriculum developer | Research Subject | external participant | paper coordinator | Graduate Student | standard~~user. | Broker |
|---|---|---|---|---|---|---|---|---|
| 23 | 62 | 13 | 3 | 2 | 3 | 1 | 1 | 6 |

| Coordinator | Offshore Technical Coordinator | Technical Co-ordinator | SCIS Resource coordinator | Content Facilitator | Developer | Officially sanctioned local developer | Programmer | Technologist |
|---|---|---|---|---|---|---|---|---|
| 60 | 41 | 2 | 5 | 1 | 8 | 2 | 2 | 1 |

| Testers. | Support and Maintenance Team representatives | Configurer | help desk staff | trainers | System Support Consultant | audiovisual unit - SLU | videoconference technicians | Supplier |
|---|---|---|---|---|---|---|---|---|
| 5 | 12 | 1 | 2 | 2 | 2 | 2 | 1 | 1 |

| IRB administrator | IRB | ISP |
|---|---|---|
| 1 | 4 | 1 |

**Figure 4.** Concepts and codes – roles for establishment episode

**Table 6.** Establishment episode - Metastructure of *GVT Formation* Process

***AUT***

Technology - Hosts service, AUTonline group pages, Lotus Notes DB, email, diagram attachment
Institutional - accepting of external registrants, slow ITS service causes delays
Individual actions - DK coordinates, FN proposes, AP advises student details
TUM - DK & TC confirm GVT & Local Team nos, advise registration, set up GVTs for FN, confirm ability to remove
Tech use - AUTOnline, group pages, manage groups, email + diagram attachment
Cultural - AUT LTs based on earlier course groups, GVTs designed to support research + teaching

***St Louis***

Technology - AUT Hosts service, AUTonline group pages, Lotus Notes DB, email, diagram attachment
Institutional - not accepting of external registrants, students enrolments unstable, no IRB approval
Individual actions - DK coordinates, FN proposes, AP advises student details
TUM - DK & TC confirm GVT & LT nos, advise registration, set up GVTs for FN, confirm ability to remove
Tech use - AUTOnline, group pages, [read mode], email + diagram attachment
Cultural - AUT LTs based on earlier course groups, GVTs designed to support research + teaching

***Uppsala***

Technology - AUT Hosts service, AUTonline group pages, Lotus Notes DB, email, diagram attachment
Institutional - supportive but delayed details, students enrolments unstable
Individual actions - DK coordinates, AP advises student details (delayed), sets up GVTs
TUM - DK & TC confirm GVT & LT nos, advise registration, confirm ability to remove, AP sets up GVTs,
Tech use - AUTOnline, group pages, manage groups, email + diagram attachment
Cultural - AUT LTs based on earlier course groups, GVTs designed to support research + teaching

These radar charts portray the degree of alignment or "collaborative technology fit" (CTF) achieved on each dimension. The scales in Figure 5 represent a continuum from zero fit to full fit, where full fit reflects an ideal situation. The ratings at this stage are based upon the authors' holistic judgement of fit on each dimension. Further work is required to calibrate the scales and to more reliably determine the degree of fit. To briefly explain the mappings above, the degree of fit appeared stronger here for the AUT site, perhaps as a consequence of it being the host site for the technology platform and for the trial coordinators, with the AUT-driven team design imposing a common cultural pattern across sites.

The **sixth** and final element of the analysis for each episode then incorporated a process of temporal bracketing, through which the evolution of the episode was charted over time. The role of TUM in how practices developed was a typical focus of this analysis, which was informed by Orlikowski [43], and typically charted in a timeline of technology, practices, activities and events (as in Figure 7). Again this more holistic analysis was informed by the intimate knowledge of the data gained from the earlier grounded forms of analysis in the episode. In many cases this form of analysis required extending the episode to the origins or destination of an evolving practice or form of technology use. The selection of an extension to the temporal bracket was typically informed by the TUM activity in focus for the episode. This temporal bracketing strategy (whether fine or broad ranging), was inherent in the selection of each episode or episode grouping, and provided a window within which realized patterns of practice could be observed.

The *establishment* episode extended over some 11 months duration as noted in Table 5, thus there is a strong temporal dimension inherent in this episode itself, and since the establishment mode of TUM activity inherently represents a phase or a ‘temporal bracket’ in a collaboration, temporal bracketing is integral to this episode. Activity levels within the episode are shown in Figure 6, which depicts the pattern of message exchanges that evolved in the major window of this episode between June and October 2004.

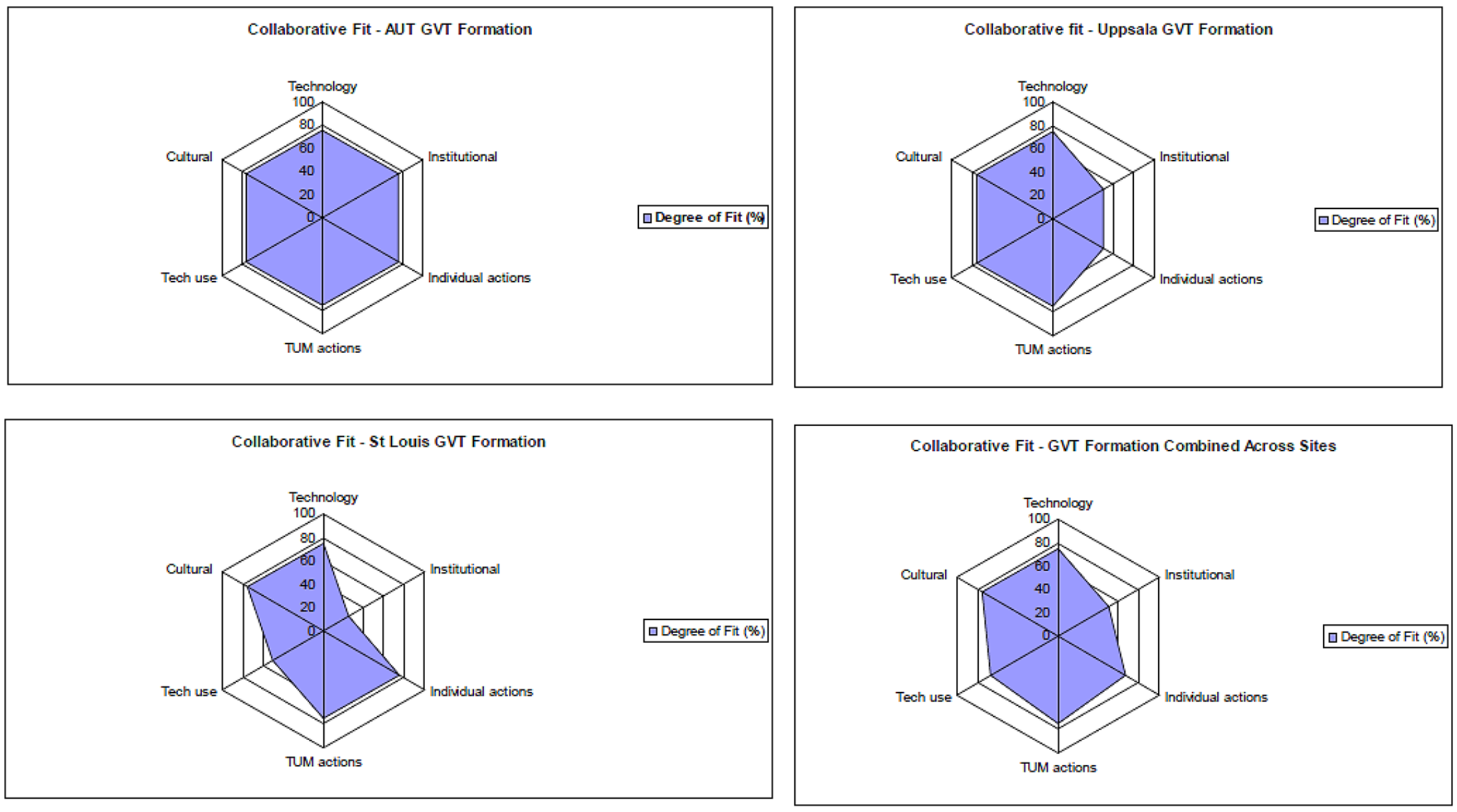

**Figure 5.** Radar charts – establishment episode - Metastructure *GVT Formation* Process

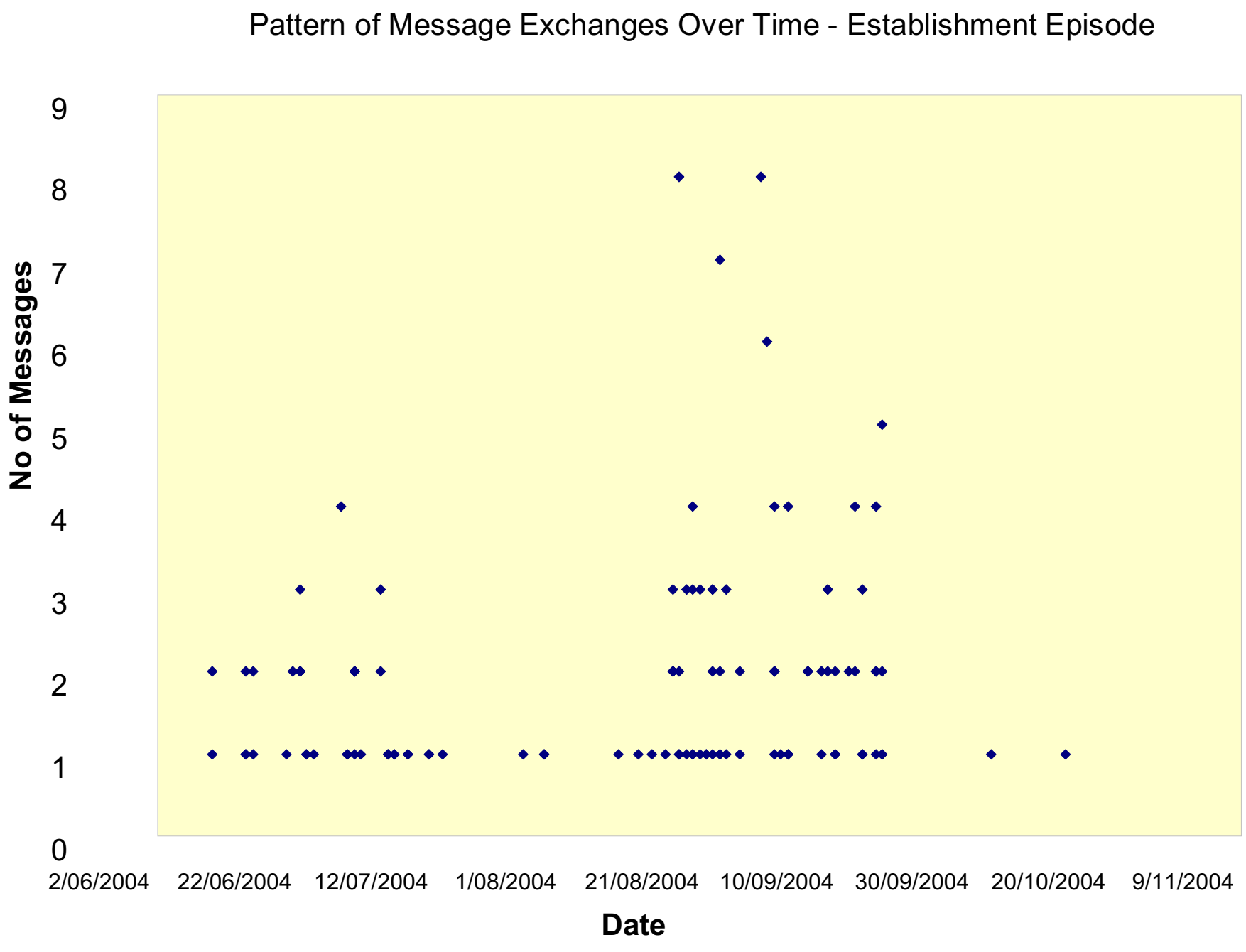

**Figure 6.** Establishment episode - Pattern of message exchanges over time

As can be seen from this scatter diagram there are three primary peaks of activity in this bracket: one in June/July some ten to twelve weeks prior to the trial; then again in mid August four weeks prior to the planned collaboration; and finally in September immediately prior to and during the first week of the collaboration.

This set of six mutually reinforcing analyses enabled an in depth encounter with the data for the episode. Both static patterns or themes and emergent patterns of practice and technology appropriation over time could be discerned. Moreover, the actors in the episode were portrayed in their institutional contexts and their detailed interactions with technology were embedded in the analysis. Two further episodes that address the remaining modes of TUM activity are now briefly summarised giving an abbreviated picture of each episode applying the same combination of analysis methods.

### *4.2. Episodic Analysis - Adjustment-Reinforcement Episode One*

This episode comprised a hybrid of the *adjustment* and *reinforcement* modes of TUM activity. The **six** elements of analysis were again carried out. The **summaries** depicted the characteristics of this much smaller episode, which included some 9 data sources, 6 actors and 2778 words, and related to a process of adjustment and reinforcement activity in readiness for phase two of the collaboration. The **appropriation analysis** indicated a predominance of 'judgement' activities where advantages of technology structures were noted [15, p.135] (e.g. the ability to create and delete demonstration entries in the collaborative database to enable testing to be carried out, and the merits of a synchronous chat session before moving into phase two of the collaboration). By contrast an example of negation (rejection of use of a technology structure) was apparent in the coordinator's decision to hold off on a 'three way phone call' to synchronise the site coordinators' views about phase two across the three locations. The **grounded theoretic analysis** identified key concepts and codes identified in the episode, then further traversed the key structurational notions of "duality of technology", "time and space" and "reflexivity of the actors" in order to unpack their operation within the episode. An example of an exploration of "time and space" in operation within the episode (where its impact was found to be particularly prominent) is briefly excerpted below:

A combination of impact of both *location* and *time* can be seen in the message below from the instructions to students:

> Please note that students from New Zealand have a two week break between the $20^{th}$ of September and the $3^{rd}$ of October. Members of GVTs are encouraged to carry on with the icebreaking process using any of the above communication channels. (DK 30/09)
>
> One *time* specific coding is that of *class schedule* in which the start and end dates for the collaboration are presented in the instructions, with specific intermediate due dates for each step of the process, and with the holiday break above clearly identified.

The **visual mapping** of selected aspects of the episode again applied radar charts to depict the operation of selected *metastructures* which portrayed the TUM activity associated with the draft instructions for the second phase of the collaboration. The patterns that emerged here demonstrated differing degrees of misalignment across sites. This misalignment resulted from several interlinked but contending dimensions. 'TUM actions' at the Auckland site proved insufficient due to time pressures and the inability of the coordinator to synchronise a three way telephone call across widely divergent time zones, so that he could confirm the phase two trial design. Therefore the instructions had to be unilaterally issued based on earlier broad agreements with the Uppsala and St Louis coordinators. The Uppsala coordinator's 'TUM actions' proved slightly contradictory when he posted an announcement to students exhorting them to arrange a synchronous chat session before phase two of the collaboration began. At the Uppsala and St Louis sites the 'technology' dimension proved problematic as the external email option within the shared platform hosted by AUT University did not support students without AUT internal email addresses. This issue arose from decisions made at an 'institutional' level at AUT, and was not initially diagnosed as a problem. Moreover there was limited support for the Uppsala coordinator and Swedish student preference for open source team-driven synchronous communication using ICQ and IRC collaborative technologies. At a 'cultural' level the semester holiday break had caused the St Louis students to lose momentum and commitment to phase two of the collaboration.

The **temporal bracketing** for this episode depicts the evolution of practices over time, both within this episode and through a logical extension. The relatively narrow focus of interest in the selected episode (synchronous technologies and AUTonline email technology use), helped in the extension of the temporal bracket. Figure 7 is presented here to illustrate the process of analysis and its depiction of specific findings from the research. The process of managing email communication and the need for remedial TUM activity is illustrated in the schematic. The following text excerpt from the study is indicative of the prior analysis of TUM activity which led to the dynamic evolution of practices depicted in Figure 7.

DK subsequently responded communicating the deficiency of AUTOnline email for contacting Swedish students, and indicating a resolution to the problem, requiring an active *TUM adjustment activity* on both the student and DK's part, since DK had individual email addresses for Swedish students.

> I'm not sure which email addresses you are using for this email to your Swedish counterparts. If you are emailing from within autonline you need to be aware that the Swedish students do not use the aut email addresses and won't get your email. You need to use their own email addresses; if you need them, let me know and I'll forward them to you. (DK 14/10)

As earlier noted…email for external students effectively went into a "black hole", so their designated home or university email addresses had to be used instead.

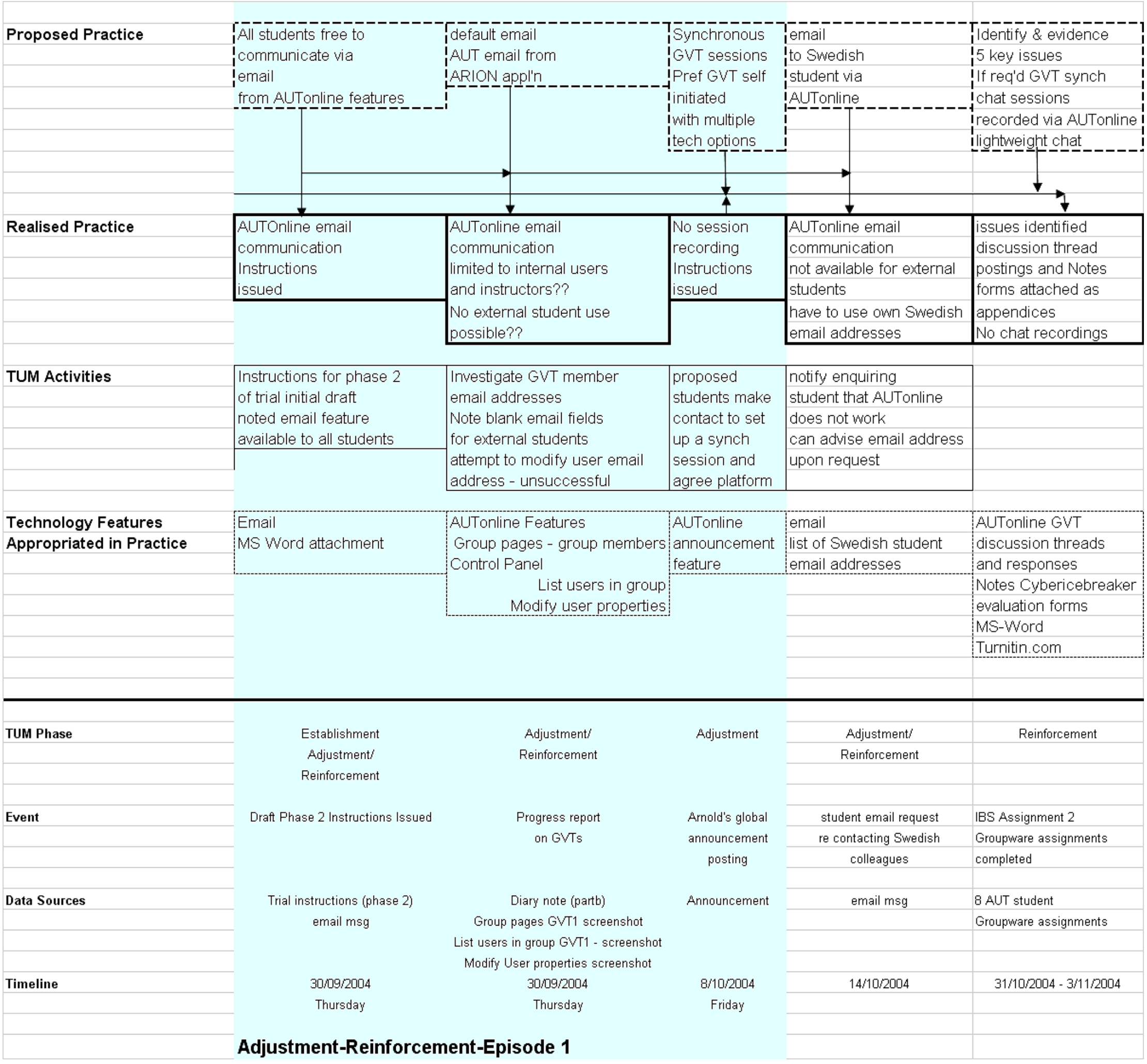

**Figure 7.** Temporal bracket: extended adjustment-reinforcement episode one - evolution over time

So through this analytical lens we were able to identify and explain misalignments or misunderstandings as causes of difficulties.

### *4.3. Episodic Analysis - Episodic Change Episode One*

This episode presented the *episodic change* mode of TUM activity. The **six** elements of analysis were again conducted. The **summaries** depicted the characteristics of this very focused episode, which included a single data source, 2 actors and 765 words, and related to a review meeting at an international conference between two coordinators of the collaboration, in which issues related to the previous collaboration were reviewed and a design for the next year's collaboration proposed. The **appropriation analysis** indicated forms of appropriation where episodic change was evidenced by moves with a 'future status' *(stating what is proposed to be done with or to establish the structure).* This latter code augmented the set of moves applied in [15] relating to setting up technology structures or features. A change in the technology structures used to support the collaboration was proposed, based upon a combination of optional communication technologies and a mandatory Lotus Notes™ database being retained for storage of key decisions, evaluations and critical research data.

The **grounded theoretic** analysis identified several discrete codes even within this small episode (e.g. technology, culture, space, time, breakdown, role etc.).

The **visual mapping** for this episode had merely tabulated the six dimensions (of Table 6 above) by TUM activity, without highlighting a specific metastructure, but still demonstrated several aspects in operation within the episode. This analysis highlighted TUM activity in the 'episodic change' mode, with 'institutional' dimensions comprising: two separate courses and assignments, site specific learning goals, and differing research ethics approval requirements at the Auckland and Uppsala sites. For the 'cultural' dimension student motivation was noted as an issue, with Uppsala student concerns to be addressed, as the focus of the collaboration upon the AUT student needs had not adequately served both cohorts. The 'technology' dimension needed addressing as usability deficiencies in the prototype Lotus Notes application had been poorly received by the Uppsala students. In the dimension of 'Technology-use mediation', activities involved the coordinators confirming mutual goals, planning the future collaboration process to suit both sites, adapting to the pre-imposed constraints of the AUT research ethics process, and agreeing the principles of the overall process, the tasks (including learning tasks) involved, and a more satisfactory technology design.

The **temporal bracketing** for this very discrete episode in itself afforded restricted scope for showing any progression of events. However widening the window of analysis into the past enabled the origins of the episode to be portrayed. In the previous year's collaboration two different technology options had been adopted (a three dimensional collaborative virtual environment cf. [13], and a two dimensional Lotus Notes prototype application). While this technology combination had not been without challenges and numerous technology breakdowns had been experienced, the combination had met with some success in motivating the Swedish students. By contrast the single platform option of a modified two dimensional Lotus Notes prototype application (with significant usability issues) had generated considerable resistance from Swedish students in the 2003 collaboration. Thus the TUM activity of 'episodic change' unpacked in this brief episode had its origins in decisions and activities occurring some months and even years before. The planning activity of the coordinators here focused upon addressing the deficiencies in the technology platform, but they could not be addressed in isolation from the other issues of course, common collaborative task yet site-specific assessment designs, and student preferences and motivations at each site.

### *4.4. Summary of Episode Profiles*

The three episodes profiled above serve to demonstrate the research process adopted here. The episodes presented have collectively covered the four modes of technology-use mediation activity and demonstrate how the sets of analysis in combination serve to build both a micro-level picture of the activity and its evolution on a larger scale. By employing the set of six mutually reinforcing strategies for episode analysis, we believe that this approach helped us to build a rich set of understandings, which were deeply grounded in the data, including the interpretations and practices of those involved, and thus provided triangulation.

While the substantive dimensions of the analyses have not been drawn out in depth here, these excerpts have been presented to demonstrate how the methods have been operationalised and specific techniques employed. Subsequent steps of the research have then combined the resulting patterns from each episode in what could be viewed as a form of cross-case analysis [17]. In that stage, patterns were compared across cases and broader conclusions were drawn through a process of interpretation, but based tightly upon the patterns and themes emerging from the data. Thus more enduring patterns of recurring concepts and practices could be derived, than might be identified in a single episode.

### *4.5. Value of the Approach and Limitations*

Before discussing the general merits of this form of analysis, it may be helpful to outline the role of the six elements applied when analysing each episode. At episode level the *first element* (a tabular summary) serves as an overview only to profile the episode for later comparison. The *second element* (narrative summary) serves as a fuller overview to summarise the episode content, both to aid the reader in comprehension and the researcher in the initial identification of codes, categories and themes. The *third element* (appropriation analysis) investigates appropriation moves to highlight TUM activities at the micro level, in order to surface patterns of technology use and TUM activity specific to that episode. The *fourth element* (grounded theoretic analysis) serves to highlight recurring codes and categories within the episode, in order to surface more general patterns specific to that episode. The *fifth element* (visual mapping) profiles selected metastructures identified from the previous analyses to show the dynamics applying within the episode. The *sixth element* (temporal bracketing) profiles the selected TUM activity in focus and the evolution of practices over the duration of the episode, extended in some cases by linking forward or backward in time to show the origin or resolution of the situation.

Yet since analysis is then conducted at the cross-episode level, these elements serve a role beyond that of the single episode. *Elements one and two* support broad identification of different types of episode (episode context, length, number of data sources and actors, dominant TUM activity modes and so on). This could be seen as somewhat similar to a description of each dataset's characteristics using both words and descriptive statistics, as would be expected in an experimental study. *Elements three and four* support comparison of general patterns across episodes and episode types (e.g. to profile dominant patterns within specific modes of TUM activity, and further to support comparison of patterns typical to differing modes - for instance the 'establishment mode' as opposed to the 'episodic change mode'). These could be seen as analogous to the application of statistical grouping or clustering techniques (e.g. principal component analysis or factor analysis), as might be adopted in an "evaluative-deductive" [47] study. *Elements five and six* support the comparison of patterns of evolution of practices for specific TUM activity modes in focus for each episode. They also enable differing profiles of highlighted metastructures to be compared across episodes and episode types. These types of analysis exemplify the more process-oriented and context-framed nature of these analytical techniques, while still enabling cross-context comparison and a degree of generalisation of findings. This we argue is a particular strength of these techniques, as complex and dynamic patterns of practice are unpacked in all their situated richness, and as they unfold. Yet these evolving patterns of practice still remain open for rigorous comparison, and much like software design patterns [19] can be reviewed for applicability or differences in application across contexts.

This analytical strategy therefore combined both discrete and more continuous elements and led to the building of broader theory. The analysis of the technology structuring processes in their context enabled the linkages between the individual actions, technology use, institutional and cultural forces to be identified and verified. Thus we could highlight to what extent technology was embedded in the actions of those engaged in the global virtual collaboration, and the role and at times limitations of actions of technology-use mediation in embedding technology in support of the collaborative process. One outcome of the study has been the development of a novel theory of Collaborative Technology Fit described in more detail elsewhere [7, 12].

One crucial point became evident from the above analysis and in particular the review of practices as they developed over time within their temporal brackets. The critical roles that time and timing had to play in this global venture were starkly apparent. Any analysis of such work without accommodating the temporal dimension would inevitably be deficient. The subtle evolution of events and their interrelationships in these episodes support the applicability of a "process research model" as opposed to a "factor research model" for this form of investigation. Newman and Robey [39] have drawn the distinction that "process models focus on sequences of events over time in order to explain how and why particular outcomes are reached", whereas a factor research model by contrast, generates "inferred processes of development". This form of temporal analysis as an example of a "process research model" focused specifically on the sequence of events and their implications as they unfolded over time.

Nonetheless the efficiency of this approach as a research method could be questioned, and the relative contributions of each form of analysis could be revisited in subsequent studies. The conclusions tended to come at the end of each episode analysis when the visual mapping and temporal bracketing had been concluded. It remains unclear to what extent the prior forms of analysis were critical to gaining the required familiarity with the data, and resulting depth of insight to develop those depictions. These separate processes demanded a quite time consuming and rigorous effort and it may well be possible to reduce these steps of analysis. Yet we remain comfortable with the appropriateness of these six reinforcing methods and accompanying techniques for an in depth initial study with a focus on inductive theory building in the complex and little understood area of technology-use mediation.

In subsequent field work validating the resulting theory of "Collaborative Technology Fit" [12], a more efficient and intuitive analysis has been performed by students engaged in a global collaboration [7]. Armed with only a guidebook developed from this work and tailored for their use, the students demonstrated the ability to apply the theory as an aid to their reflection upon their global collaboration process. As a first step, they were independently able to identify both 'episodes of interest' and associated 'metastructures'. The first metastructure identified was the collaborative technology platform '*ClockingIT*', a cloud computing application which had been adopted to support the collaboration cf. [7]. The software was designed to support project management and time tracking through features including "chat rooms, instant messaging, a built in wiki and discussion boards" [7]. In addition to the *technology* metastructure of '*ClockingIT*' they identified a 'team leader meeting' as an *institutional* metastructure. Using a fairly loose approach they appeared readily able to identify relevant aspects of each dimension of Collaborative Technology Fit (CTF) and these were then mapped using the tabulation and radar chart approach described above in the fifth element of the TUMAST episode analysis method. They did gain insights from these analyses which aided their reflection. For instance:

> *"students performed a CTF-analysis on how a team leader meeting was carried out, from planning it, through to the actual meeting taking place, ending with the planning of the next meeting. During the session, it was recognized that due to differences in technical equipment, team leaders from Sweden and from America experienced the conference call part of the meeting differently. In Sweden, the students had access to a well-equipped videoconference room, while the American students used their own PC's. This resulted in non-optimal communication, due to poor sound quality and sometimes distracting surroundings. This CTF-session thus resulted in a discovery of previously unknown reasons for communication difficulties"* [8].

In addition they noted some insensitivity to *cultural* issues relating to the choice of a time for the meeting to best suit the Swedish students, but which demanded an early rising for their American counterparts [7]. Further they were able to quite consciously explicate the processes of *technology-use mediation* they had engaged in when setting up the team leader meeting, by noting specific actions such as: the Team coordinator creating a task in *ClockingIT*, the coordinator booking the [VideoConference] Room, an agenda being created on the wiki, audio equipment being hooked up and so on [7].

So it appears that the results from the demanding form of research outlined in this paper may potentially be translated into suitable instruments and, in a reduced form, applied more pragmatically and efficiently in other practice settings. The extent then to which all six elements of the episode analysis are crucial for drawing out the critical patterns from each episode remains to be further investigated. While they proved useful and effective for us in the early study described here, we make no claim for this being an optimal form of analysis. In combination these six elements of analysis have provided an opportunity for very deep acquaintance with the data, sufficient to enable theory building. Follow-up studies by contrast might use fewer of these elements, cut down the depth of the analysis, or link the elements in differing sequences. A related question is how the appropriation analysis may be better conducted. The decision to code individual email message segments rather than email sequences may have disaggregated these moves too much, and it has since been suggested [45] that linking from the broader picture of the temporal analysis to the sequences of appropriation moves as evolving interactions by the parties to the episode may provide different, and perhaps greater, insight.

## *4.6. Assuring Validity*

As stated in Section 2 of this paper, much software engineering research draws on the positivist or natural science approach, and sensibly utilizes tests of validity common within that research paradigm. Investigators are often concerned about the construct, criterion and content validity of measurement instruments and the internal and external validity of their experimental or perhaps quasi-experimental designs [14, p.125]. Commonly researchers within this paradigm, have an interest in causal relationships and predicting outcomes in given situations. The research is frequently guided by what Gregor [25, p.625] has termed a "Type III: Theory for Predicting".

As earlier noted, this study, in contrast, aimed to produce (in Gregor's terms) a "theory for explaining", with a primary focus on how and why TUM occurs in a global virtual team context. When judging the validity of differing forms of research Creswell and Miller [14] have argued that "the choice of validity procedures is governed by two perspectives: the lens researchers choose to validate their studies and researchers' paradigm assumptions" (p. 124).

Table 7 presents the three proposed lenses for validating studies (the lens of the researcher, the lens of the study participants and the lens of those external to the study) within three separate paradigms of 'qualitative' research. A discussion of research paradigms in relation to this study of the actions and interactions of GVTs has been presented above – the work here is broadly consistent with the *constructivist* paradigm outlined in Table 7, while sharing some elements of the *critical* paradigm. Thus the procedures for assessMcLeod, MacDonell and Doolin (2011) EMSEing validity differ from those appropriate to the *postpositivist* paradigm, the approach that would be more familiar to those schooled in the objective or natural science tradition.

**Table 7.** Validity procedures for qualitative research (ex. [14, p.126])

| Paradigm assumption/Lens | Postpositivist or Systematic Paradigm | Constructivist Paradigm | Critical Paradigm |
|---|---|---|---|
| Lens of the Researcher | Triangulation | Disconfirming evidence | Researcher reflexivity |
| Lens of Study Participants | Member checking | Prolonged engagement in the field | Collaboration |
| Lens of People External to the Study (Reviewers, Readers) | The audit trail | Thick, rich description | Peer debriefing |

For Creswell and Miller [14] *credibility* is the key yardstick for judging qualitative research (p. 124):

> *we define validity as how accurately the account represents participants' realities of the social phenomena and is credible to them... Procedures for validity include those strategies used by researchers to establish the credibility of their study.*

**Table 8.** Assessment of research quality and rigour in this interpretive field study

| **Evaluating Interpretive Field Studies** | |
|---|---|
| **Principle** | **This Study** |
| 1. The fundamental principle of the hermeneutic circle | Integral to the multi level episodic analysis within the study. The four elements of structurational analysis oscillate between micro level data and macro level context. Analysis proceeds from appropriation moves to duality of technology. The multiple levels of culture explored in the analysis of episodes move consciously from the individual to the global level. |
| 2. The principle of contextualization | The situated nature of the research, historicity of the research programme and key role of context acknowledged and explored. Each episode set in context through the structurational analysis, and temporal bracketing processes highlight key events and meetings. |
| 3. The principle of interaction between the researcher and the subjects | Researcher role, and motivation for the research outlined. Processes for participation and research design explicit. Episodic analysis makes explicit researcher and "subjects" interactions through dialogues and reflections. |
| 4. The principle of abstraction and generalization | The study has applied a number of different frameworks and theories to support the analysis, draw conclusions and suggest areas for further work. The TUMAST (Technology-use mediated AST) and CTF (collaborative technology fit) frameworks directly result from this study. |
| 5. The principle of dialogical reasoning | *"Requires sensitivity to possible contradictions between the theoretical preconceptions guiding the research design and actual findings ("the story which the data tell") with subsequent cycles of revision"* [34]. Particular pre- and misconceptions conceptual and methodological have been explored with illustrations of how they have been instrumental in directing the work. The evolution from AST to TUMAST downplaying broad analysis of activities, and to CTF with multilayered models of culture provide some relevant examples. |
| 6. The principle of multiple interpretations | The study draws together diverse forms of data, and differing voices of the actors. The distinctions between these views are consciously addressed through multi dimensional forms of analysis which triangulate across perspectives. |
| 7. The principle of suspicion | The principle of suspicion came periodically to the fore in the work. The review of constraints imposed by the context, institutional and global forces, e.g. institutional security regimes and ethical approval processes, frequently demanded a broader critique. |

In a similar vein Klein and Myers [34] have delineated seven principles for evaluating interpretive field studies. Their framework provides one broad set of principles by which to assess whether the research adheres to the tenets of the interpretive paradigm and its measures of research quality. Consistent with the reflective principles of action research, Melrose [38] has recommended: "Self-reflection on the [investigator's] learning and progress as an action researcher and/or practitioner is an important part of the [study]". Accordingly an evaluation of the field study against the seven principles of Klein and Myers was conducted to reflect upon how they had been realized [12], summarised in Table 8.

The strategies in Tables 7 and 8 then, represent procedures for judging validity in qualitative and 'interpretive' research studies. Unlike the statistical tests of the positivist science model, credibility and consistency of the account and a conscious and critical reflection upon the elements of the work are key considerations for interpretive studies.

# 5. CONCLUSIONS AND FUTURE WORK

Glass et al. [23] initially studied and categorised a set of publications in representative journals for the computing disciplines of Computer Science, Software Engineering and Information Systems over a five year period (1995-1999). Their tabulation of research methods showed a huge paucity in the software engineering field of the type of research described in this paper – for instance, action research 0%; field study < 1%; grounded theory < 1%; hermeneutics < 1% (and structuration theory did not receive a mention). Subsequent studies [47, 30, 52] have continued to report the relative infrequency of evaluative studies adopting an interpretivist paradigm and methods such as those above. Yet some growth in the use of the case study research method, although largely classified as "descriptive" in research approach, has been observed [30]. As identified in Table 2 above the pattern seems to have been changing over time, especially for global software development research, and as Dittrich et al. [16, p.534] have observed, recent trends have seen more researchers using "qualitative data" and "trying to understand more deeply the social side of software engineering".

Nonetheless it is our argument here that the increasingly global setting in which software engineering practice is being conducted demands a more extensive repertoire of research approaches and techniques than is generally used at present, if we are to better understand the complexity of current practice, let alone provide guidance for further innovation. From a similar perspective Glass et al. [24, p.505] criticised the "decoupling between research in the computing field and the state of the practice of the field", and posed the question whether slow rates of transfer from research to practice are a function of the "irrelevance of the research or the intransigence of the practitioners". Of more direct relevance to this paper, they also questioned "the narrowness of the choices made by SE researchers" and suggested that benefits might accrue from "broadening research approaches and methods", with "case and field studies" potentially providing "richer and more valuable findings" (p. 504).

Thus we argue that researching the nature of global virtual teamwork, and such phenomena as *technology-use mediation* in particular, require us to adopt new methods, capable of accommodating the inherent complexity of such situated work, work which is of critical importance to "the practice of the field". It is our contention that continued excessive reliance upon 'factor' models of research will generate laboratory-derived and simplistic models divorced from reality, or models more complex than useful, to the point of depicting "causal arrows flying in every direction", a caution given by Fulk et al. [18, p.126] in similar circumstances, referring to the "social influence model".

The study reported here has adopted an interpretive perspective, but one that requires a very deep engagement with the data, from which empirical findings can be drawn and theory built. The complementary set of research methods and data analysis techniques have aided in developing an understanding of the operation of *technology-use mediation in a global context*, incorporating elements of the cultural and the interaction between the technology, individual and institutional dimensions. It has proven capable of tracing the inherent complexities of patterns of practice as they have evolved over time, phenomena which defy reductionist and solely quantitative analyses.

We have argued here that research applying an alternative worldview, namely the interpretive perspective, can make significant contributions to our understanding of globally distributed teams. After all, these teams are populated by people embedded in cultural settings, with feelings and emotions that drive behaviours and inform practices, which may introduce risk and impact performance. The complex range of human and social phenomena warranting investigation demands an expanded repertoire of research tools and techniques. For instance, if (as found in this study) 'socio-emotional' dimensions are critical to the effective performance of distributed software teams, it will be vital to have interpretive research tools to better understand the feelings, perceptions and motivations of participants in these settings. This expanded research repertoire may also demand an extension of worldview to accept that an 'interpreted' set of understandings can offer an equally legitimate set of findings to that furnished by the more traditional 'objective science' model of software engineering research approaches.

With more specific reference to this study, the insights into global teamwork have so far been gained in educational settings, where virtual teams of professional educators, researchers, developers and technical administrators have been profiled, along with the work of global student teams. Therefore the next domain for intensive field work will be in more commercial settings, with both small and large scale projects and organizations. Through the application of the research techniques outlined above, it is believed that the work will prove applicable in commercial global software development team contexts. In the first instance we see it applying most probably in the more loosely managed collaborations and the smaller software company environment, or perhaps for larger projects in the more interactive phases of development such as requirements engineering. We believe that such techniques could make a contribution to our understanding and ability to better manage collaborative technologies and the innate risks in distributed software development.

Thus having in the course of this study developed a "theory for explaining" [25], the next steps will involve refining the theory, together with the techniques developed from the research, in commercial practice settings. Subsequently or perhaps in parallel, there is scope for research that develops a "theory for predicting". We conjecture that certain typical patterns in the mediation of technology-use may pre-exist or emerge within global virtual team settings, and moreover that positive patterns (if identified early) may be reinforced and negative patterns may be avoided, or at least have their impacts reduced. A current limitation of the theoretical model derived from this study is that it enables us with considerable effort to perform post-hoc analysis of a situation and then *lament* an unsatisfactory outcome. A clearly preferable outcome would be to be able to predict successful outcomes from the outset and thus *prevent* rather than *lament*.

Continuing such research in this interpretive vein would see further development of the methods outlined above through their application within additional field settings. Designing a suitable programme of action research to test specific interventions through a series of planned field studies, would be one strategy for developing results which could be generalized across multiple contexts. Kock for instance has advocated the use of successive action research cycles to extend both "research scope" and resulting "model generality", arguing that "progress through iterations allows the researcher to gradually broaden the research scope and in consequence add generality to the research findings" [35]. Such a research programme in the interpretive paradigm would echo that of a series or family of experimental studies conducted in the positivist tradition, but have the added benefit of being strongly embedded within a realistic context of practice.

In conclusion then the research perspective and methods outlined here should be considered part of a toolkit of approaches and methods. The interpretive methods outlined in this paper constitute one set of techniques only and other forms of enquiry are equally valid in their relevant contexts. We believe that the choice of method is best driven by the research goals and questions and the most suitable means of addressing them. However we need to be aware that we all undertake our enquiry with a value orientation, whether that is visible or not, and like a form of blindness the lenses we wear may limit our ability to see. This paper suggests that we experiment with different research methodologies and methods so that we have a more rounded repertoire to address the most important and relevant issues in global software development research, with the forms of rigour that suit the chosen approach.

## ACKNOWLEDGEMENTS

The contributions of collaborative colleagues (Mats Daniels and Arnold Pears at Uppsala University; Fred Niederman at St Louis University; Diana Kassabova and Kitty Ko at AUT University), students from Uppsala, St Louis and AUT Universities and the Late John Hughes as co-supervisor of the study are all gratefully acknowledged. We also thank the anonymous reviewers and the editors of the special issue for their insightful and constructive feedback which has greatly improved this paper.